\documentclass[twocolumn,pre,aps,showpacs,amsmath,amssymb]{revtex4-1}
\usepackage{graphicx}
\usepackage{bm}
\usepackage{epstopdf}
\usepackage{epsfig}
\usepackage{dsfont}
\begin{document}

\title{Elements of a new approach to time in Quantum Mechanics}
\author{Eduardo O. Dias}
\email[]{corresponding author: eduardodias@df.ufpe.br}
\author{Fernando Parisio}
\email[]{parisio@df.ufpe.br}
\affiliation{Departamento de
F\'{\i}sica, Universidade Federal de Pernambuco, Recife, Pernambuco
50670-901 Brazil}


\begin{abstract}
In this work we present a re-evaluation of the
concept of time in non-relativistic quantum theory. We suggest a formalism in which
time is changed into the status of an operator, and where
expectation values of observables and the state of a quantum system
are reworked. This approach leads us to an additional concept, given
by a temporal probability distribution associated with the actual measurement
of an observable.
\end{abstract}
\maketitle

\section{Introduction}
\label{intro}

In Schr\"odinger quantum mechanics there is a clear
asymmetry between time and space. Time is treated as
a continuous parameter that can be chosen with arbitrary precision
and employed to label the solution of the wave
equation. A character of infinite divisibility of time is taken for granted.
In contrast, the position of a particle is treated as an
operator, and therefore, the knowledge of its value becomes inherently
probabilistic. It is quite common to find the reasoning that this
asymmetry is due to the non-relativistic character of the Schr$\ddot
{\rm o}$dinger equation. Although partially correct, this argument is
insufficient to justify all the disparity between space and time in the
formalism of quantum mechanics. An illustration is as follows:
the squared modulus of the wave function provides the probability,
in a position measurement, of finding the particle between $x$ and
$x + {\rm d}x$ exactly at the instant $t$. Would it not be
more reasonable and symmetrical, even in the non-relativistic domain, to
ask about the probability of a particle to be measured between $x$
and $x + {\rm d}x$ at an instant between $t$ and $t + {\rm d}t$?

The first goal of this manuscript is to extend certain concepts in
quantum mechanics and show that the asymmetry between $t$ and $x$
is, in fact, not entirely due to the lack of Lorentz covariance of
the theory. Firstly, the study proposed here is motivated by the
uncertainty relation between energy and time. The inequality $\Delta
E\Delta T \geq \hbar / 2$ has more diverse interpretations than
those involving two canonically conjugate observables. Consider the
case of a stationary state, for which we can determine with
certainty the energy of the system and, hence, $\Delta E = 0$.
Consequently, we should have ``$\Delta T = \infty$'', so that a
plausible interpretation of $\Delta T$ would be the lifetime of the
state. Alternatively, in his book ``Quantum Mechanics''~\cite{Grif},
D. J. Griffiths presents a derivation of the energy-time uncertainty
relation, in which, an observable $\hat A$ is measured and $\Delta
T$ plays the role of the time required for $\langle A \rangle$ to
vary by a standard deviation of $\hat A$. Therefore, by D. J.
Griffiths ``$\Delta T = \infty$'' indicates that the expectation
value of the associated observable is constant in time. If the
operator $\hat{A}$ and the Hamiltonian do not commute, these two
interpretations not necessarily coincide.

It is possible to think of quantum dynamics as a combination of two
quite disparate processes: a unitary evolution, and the collapse
caused by measurements. Given a $|\psi (t)\rangle$ satisfying the
first process, in quantum theory we can select $t$ with arbitrary
precision to analyse possible results of the second process. The
fact is that this arbitrary temporal choice does not take into
account the fundamental aspect that any information about a quantum
system is only obtained, in practice, by a measurement that has an
intrinsic temporal randomness. A second goal of this work is to
review the concepts of quantum mechanics associated with this
analysis, taking into consideration the fact that we do not have an
infinitely precise knowledge of the time in which we ``observe'' the
system. In order to do this, we abandon the idea that one can
describe the state of a system at a given instant of time. In this
case, we can also define a time operator that provides a starting
point for a more systematic approach to the concept of time in
non-relativistic quantum mechanics.

Operators related to time have already been addressed in several different
contexts~\cite{Wigner,Kijowski,Muga,Werner,Hartle,Marinov,Grot,Kumar,toperator,time}.
Among these approaches, an important problem is related to the
arrival time of a particle in an apparatus that is
spatially localized. In this scenario, a time operator is defined so
that the relation $[{\hat T}, {\hat H}] = i\hbar$ is satisfied, and
the main objective is to obtain the probability distribution for the
time of arrival at the detector, quite a useful
concept, especially from an experimental point of view. This idea
brought significant academic interest and numerous case studies,
as for example in quantum tunneling~\cite{Marinov}.
We will see later that the physical motivation and the way the
operator associated with the arrival time is defined is quite different and,
indeed, complementary with respect the one proposed in this
work. Some of our results on this topic are discussed in next
section.

We should point out that a very recent work entitled ``Quantum
time''~\cite{time} proposed a new, consistent way to understand time
in quantum mechanics. Although we followed a distinct path, our
equation (10) is equivalent to Eq.~(23) of Ref.~\cite{time}. The
study carried out in Ref.~\cite{time} is based in a concept,
originally proposed by Page and Wootters~\cite{Wooters}, of the
conditional character of the quantum state relative to a Hamiltonian
system defined as a clock. Our approach embraces the same
conditional interpretation for the solution of the Schr$\ddot {\rm
o}$dinger equation by using quite different arguments and physical
motivation.

\section{Space-time symmetry and the role of time in Quantum Mechanics}
\label{derivation}

Some symmetries between time and
position, often concealed by the standard presentation of the theory, can be found in
operators and equations of non-relativistic quantum mechanics. For instance:
\begin{equation}
\label{evolve1} {\hat H} {\hat U}_t(t,t')=i\hbar \frac{\rm d}{{\rm
d}t}{\hat U}_t(t,t')
\end{equation}
and
\begin{equation}
\label{evolve2}{\hat P} {\hat U}_x(x,x')=i\hbar \frac{\rm d}{{\rm
d}x}{\hat U}_x(x,x'),
\end{equation}
where ${\hat H}$ is the Hamiltonian of the system, ${\hat U} _t$ is
the time evolution operator (or temporal translation operator),
$\hat P$ is the momentum operator, and ${\hat U} _x$ is the spatial
translation operator. By acting the ket $|\psi(t')\rangle$ (solution
of the Schr$\ddot {\rm o}$dinger equation) on the the right hand
side of Eq.~(\ref{evolve1}), and acting $|x'\rangle$ in the same way
in Eq.~(\ref{evolve2}), we obtain
\begin{equation}
\label{evolve3} {\hat H} |\psi(t)\rangle=i\hbar \frac{\rm d}{{\rm
d}t}|\psi(t)\rangle
\end{equation}
and
\begin{equation}
\label{evolve4}{\hat P}|x\rangle=i\hbar \frac{\rm d}{{\rm
d}x}|x\rangle,
\end{equation}
where the interplay
between pairs $\hat H$ and $\hat P$, ${\hat U}_t$
and ${\hat U}_x$, and {\it also} between $|\psi(t)\rangle$ and
$|x\rangle$ is clear. Because of the similarity between Eqs.~(\ref{evolve3})
and (\ref{evolve4}), we are compelled to tackle the delicate problem
of defining a time operator $\hat T$ whose eigenvalues correspond to the
very solutions of the time-dependent Schr$\ddot {\rm o}$dinger equation for each instant
of time. So, defining $|\psi(t)\rangle\equiv|t\rangle$, where ${\hat
T}|t\rangle=t|t\rangle$, we have the following relations
\begin{equation}
\label{evolve5} {\hat H} |t\rangle=i\hbar \frac{\rm d}{{\rm
d}t}|t\rangle~{\rm ,}~~{\hat P}|x\rangle=i\hbar \frac{\rm d}{{\rm
d}x}|x\rangle,
\end{equation}
and
\begin{equation}
\label{evolve6} {\hat T} |t\rangle=t~|t\rangle~{\rm ,}~~{\hat
X}|x\rangle=x|x\rangle.
\end{equation}
Note that $x$ and $t$ are placed in a quite symmetrical
framework. However, one
can easily note that $\hat T$ is non-Hermitian, whereas
$\hat X=\hat{X}^{\dagger}$ . This has been one of the biggest concerns regarding the
definition of a time operator.

It is worth to point out that a similar operator was defined in
Ref.~\cite{toperator} through a distinct physical reasoning.
In~\cite{toperator} a picture of quantum mechanics is assumed in
which the representation vectors evolve in time and state vectors
and operators are static, which lead to very different conclusions
in comparison to ours.

The approach described above seems to provide interesting physical
consequences, while it does not affect the existing results in quantum
mechanics. Consider, for example, a Hamiltonian with an explicit
time dependence. According to our suggestion, we should
replace the parameter $t$ by the operator $\hat T$. However, in the
Schr$\ddot {\rm o}$dinger equation, as the Hamiltonian acts in the
eigenvector of $\hat T$, $|t\rangle$, time becomes a scalar quantity again, and,
thus, nothing is changed in comparison with standard quantum theory.
Another important consequence of the existence of this operator is
that we can easily obtain the relation $[{\hat T}, {\hat H}] =
i\hbar$. However, by looking at the dispersion of $\hat{T}$ a
problem arises, which leads us to rework some concepts as follows.

The calculation of the expectation value of the time operator for a
physical system whose ket state is $|\psi(t)\rangle$ reads
\begin{equation}
\label{average1} \langle \psi(t)|{\hat T}|\psi(t) \rangle= \langle
t|{\hat T}|t \rangle=t~\langle t|t \rangle= t,
\end{equation}
which implies that the temporal uncertainty vanishes, $\Delta
T=\langle T^2\rangle - \langle T \rangle^2=0$, regardless of the
details of the physical system considered. As discussed earlier, the reason why
$\Delta T = 0$ is the usual approach of quantum mechanics
that considers time as a parameter that can be chosen arbitrarily.
Moreover, this result, conceptually unsatisfactory, could not take
us to the Heisenberg inequality via first principles. Therefore,
we are lead to a redefinition of quantum expectation values.
Let $\hat A$ be an observable, the modified expectation
value of $\hat A$ reads
\begin{equation}
\label{average_new} \langle\langle{\hat A}\rangle\rangle
\equiv\int_{0}^{\infty} f(t) \langle {\hat A} \rangle {\rm d}t.
\end{equation}
Eq.~(\ref{average_new}) corresponds to a time average of the usual
expectation value of quantum mechanics $\langle \hat{A} \rangle$
weighted by the function $f(t)$, which satisfies the condition
$\int_{0}^{\infty} f(t) {\rm d}t=1$. The function $f(t)$ is
interpreted as the temporal probability distribution for
the wave function to collapse once the system is under measurement.
We stress that the formalism described from this point on is
intended to be valid only in the temporal window where the collapse
can potentially occur, that is, after the arrival time mentioned in the
introduction has elapsed. A similar interpretation appears in a distinct, more specific
context, in studies on physical collapse dynamics~\cite{Parisio1,Parisio2}.

Note, in particular, that for ${\hat A} = {\hat T}$, we have
\begin{equation}
\label{average_time} \langle\langle{\hat
T}\rangle\rangle=\int_{0}^{\infty} f(t) t~{\rm d}t,
\end{equation}
where a clear difference from the expectation value given by
Eq.~(\ref{average1}) is observed. By using the interpretation of
$f(t)$ given earlier, Eq.~(\ref{average_time}) yields the average
time for the collapse to occur in a set of measurements over an
ensemble of equally prepared physical systems. Here a key concern
arises: what kind of quantum state would be consistent with the
expression of Eq.~(\ref{average_new})? To answer this question, we
redefine the quantum state of the system using the density operator
formalism. The quantum state of a system {\it under measurement}, with the
solution of the Schr$\ddot {\rm o}$dinger equation given by
$|t\rangle$, is defined as
\begin{equation}
\label{state} {\hat \Omega}=\int_{0}^{\infty} f(t) |t\rangle\langle
t| {\rm d}t.
\end{equation}
Note that the state of Eq.~(\ref{state}) does not correspond to a
representation of the usual density operator of quantum mechanics
obtained by using the states {$|\psi(t)\rangle =|t\rangle$}. In
fact, these states are not even a basis for the Hilbert space.
Eq.~(\ref{state}) represents a new quantum state given by the sum of
the Schr$\ddot {\rm o}$dinger equation solutions for each instant of
time, weighted by the function $f(t)$. Note that Eq.~(\ref{state})
does not consider the time as a parameter whose value can be
arbitrarily chosen. One can recover the standart theory by setting
$f(t)=\delta(t-t')$. As mentioned in the introduction, one can
obtain the state~(\ref{state}) through the state defined in Eq.~(23)
of Ref.~\cite{time}. In order to do this, one traces out the clock
space, defined by the authors as an ancillary system.

Note that the quantum state proposed in this work corresponds to a
temporal convolution of solutions to the Schr\"odinger equation,
being independent of space and time. This approach is markedly
distinct from that of standard quantum mechanics, where the state is
described for a particular time. Furthermore, every physical theory
should describe the results of observations which, in some instance,
correspond to classical concepts that make our comprehension viable.
The suggestions presented here produce probabilities that an
observation will be materialised in a given time window and in a
given region of space. In this way, to infer about the state of a
particle at a given time $t$, makes as much sense as to ask about
the state of that particle in a given position $x$. Therefore, the
physical description of a quantum state in a fixed time that
instantly changes due to a measurement is incompatible with the
description presented here. By the same token, consider the
probabilistic process of spontaneous decay of an atom. It is
described by a superposition
$(|e\rangle|0\rangle+|g\rangle|1\rangle)/\sqrt{2}$ that contemplates
the existence and the non-existence of a photon. Whether or not the
photon exists immediately before the measurement (whether or not the
cat is dead or alive) becomes an ill-posed question in the presented
framework.

With these ideas in mind, consider a Stern-Gerlach experiment in
which one of the detectors is removed. If no click is observed after
a certain period of time, the collapse of the the quantum system
occurs so that one can conclude that the atom ran through the path
where the detector was removed. In these circumstances, one can ask
about either what is the state of the particle in a given instant of
time or, in other words, what is the moment at which the
Schr\"odinger evolution is no longer valid and the instantaneous
change of the state takes place. However, notice that
since~(\ref{state}) is a temporal superposition of Schr\"odinger
solutions, the very validity of this kind of question becomes
disputable.

According to one of the mantras of the Copenhagen interpretation, the idea that the position of particle
becomes a physical reality only when a measurement is made.
It is a tenable position to expect that the concept of time should  emerge in the same way.
Successive observations would give us the notions of space and time that we classically experience.

It is important to note that all usual properties of
the density operator are kept for the state~(\ref{state}), for
example
\begin{equation}
\label{prop1} {\rm Tr}~{\hat \Omega} = {\mathds I},
\end{equation}
and
\begin{equation}
\label{prop2} {\rm Tr} ({\hat \Omega} {\hat A}) =\langle\langle{\hat
A}\rangle\rangle=  \int_{0}^{\infty} f(t) \langle {\hat A} \rangle~
{\rm d}t,
\end{equation}
where ${\mathds I}$  is the identity operator. In this
formalism, the probability of finding a particle between $x$ and $x
+ {\rm d}x$ at an infinitely precise instant of time $t$ must be zero.
However, the probability of finding the particle between $x$ and $x
+ {\rm d}x$ in the the interval of time $t$ and $t + {\rm d}t$ is given by
\begin{eqnarray}
\nonumber \label{prob1} p(x,t)~{\rm d}x {\rm d}t &=& f(t) |\langle
x|t \rangle|^2~{\rm d}x {\rm d}t\\ &=& f(t) |\psi(x,t)|^2 ~{\rm d}x
{\rm d}t.
\end{eqnarray}
This new definition of quantum probability can be obtained from
Eq.~(\ref{state}) by calculating
\begin{equation}
\label{prob2} {\rm Tr} ({\hat \Omega} |a\rangle\langle a|) =
\int_{0}^{\infty} f(t) |\langle a|t \rangle|^2{\rm d}t,
\end{equation}
and by assuming $|a\rangle =|x\rangle$.

By inspecting Eq.~\ref{prob1}, we realize the conditional character
of the probability density $|\psi(x,t)|^2$. According to the Bayes
rule, one can assert that the probability of finding the particle
between $x$ and $x + {\rm d}x$ in the the interval of time $t$ and
$t + {\rm d}t$ is equal to the probability of finding the particle
between $x$ and $x+{\rm d}x$ \emph{given that} the measurement
occurred in the interval $[t,t+{\rm d}t$ times the probability for
the system to be measured between the instants $t$ and $t+{\rm d}t$:
\begin{eqnarray}
\nonumber \label{cond}p(x,t)~{\rm d}x {\rm d}t &=& P(x|t) P(t)~{\rm
d}x {\rm d}t \\
&=& |\psi(x,t)|^2 f(t) ~{\rm d}x {\rm d}t.
\end{eqnarray}
In other context, a conditional property of the wave function has
already been addressed in Ref.~\cite{Wooters} and reinforced in
Ref.~\cite{time}. Notice that $|\psi(x,t)|^2$ is interpreted now as
the probability density of finding the particle at the position $x$
given that the measurement occurred at the time $t$, so that perhaps
we should express the wave function with a more provocative
notation, such as $\psi(x|t)$. It is essential to note that by
strictly following Bayes rule, one must not assume that the function
$f(t)$ can be obtained by the knowledge of
$|\psi(t)\rangle=|t\rangle$, the solution of the Schr\"odinger
equation. It would correspond to essentially new information
necessary to express the full state of the system as in Eq.
(\ref{state}). In addition, by the symmetry of Bayes rule, in
principle, we can express the probability of finding a particle in
$[x,x+{\rm d}x]$ and $[t,t+{\rm d}t]$ as
\begin{eqnarray}
\nonumber
\label{cond2}
p(x,t)~{\rm d}x {\rm d}t &=& P(t|x) P(x)~{\rm
d}x {\rm d}t\\ &=&|\phi(t|x)|^2 g(x) ~{\rm d}x {\rm d}t,
\end{eqnarray}
where $|\phi(t|x)|^2$ corresponds to the probability density of
finding the particle at the instant $t$ \emph{given that} the result
of the position measurement is $x$, and $g(x)$ is the density
probability for the system to be measured at $x$.

We stress that the average time given by Eq.~(\ref{average_time}) has a
different meaning from that of the expectation value of
the ``arrival time'' operator, discussed earlier. The calculation of
Eq.~(\ref{average_time}) provides an average time for the quantum state to collapse
which, in general, is much shorter than the average
time of arrival of a particle in an actual experimental situation. In our
case, we treat the system as if it had already ``arrived'' at the
detector and we do not know at what time the collapse should occur.

\section{Perspectives and conclusions}
\label{conclude}

As a perspective, a first step would be to obtain constraints on
$f(t)$ or even to propose an analytical form for this distribution,
and to test its predictions in measurement processes. Furthermore,
we lack a deeper understanding of the unconventional aspects, which
naturally arise, such as the non-Hermiticity of the time operator.
Several other fundamental questions arise due to these new concepts:
is the function $f(t)$ a property of the quantum system itself or
should it be described considering the physical characteristics of
the measuring apparatus? As $f(t)$ refers to the probability of
collapse, does this temporal distribution have universal properties
which are independent of the observable that is being measured? It
would be also interesting to apply this new approach to physical
situations widely discussed in the literature, for example,
tunneling. Besides, it is important to make a careful analysis of
these concepts in comparison to previous studies related to time
operators~\cite{Wigner,Kijowski,Muga,Werner,Hartle,Marinov,Grot,Kumar}.
We believe that further developments in the presented model and its
experimental investigation may help us to get a better understanding
of the mechanism of quantum collapse~\cite{Parisio1,Parisio2} and of
the role of time in quantum theory. Comments and suggestions on
these preliminary ideas are very welcome.

\begin{acknowledgments}
E. O. D. acknowledges financial support from FACEPE through its PPP
Project No. APQ-0800-1.05/14. F. P. thanks financial support from
CNPq through the Instituto Nacional de Ci\^encia e
Tecnologia~-~Informa\c{c}\~ao Qu\^antica (INCT-IQ).
\end{acknowledgments}


\begin{thebibliography}{99}

\bibitem{Grif}D. J. Griffiths, \emph{Introduction to Quantum Mechanics}, 2nd Edition, Addison-Wesley (2005).
\bibitem{Eisberg} R. Eisberg and R. Resnick, \emph{Quantum Physics of Atoms, Molecules, Solids, Nuclei, and Particles}, John Wiley \& Sons (1985).
\bibitem{Wigner} E. P. Wigner, Phys. Rev. 98, 145 (1955).
\bibitem{Kijowski} J. Kijowski, Rep. Math. Phys. 6, 361 (1974).
\bibitem{Muga} J. S. Muga, S. Brouard, and D. Macías, Ann. Phys. (N.Y.) 240, 351 (1995).
\bibitem{Werner} R. Werner, J. Math. Phys. 27, 793 (1986).
\bibitem{Hartle} J. B. Hartle, in \emph{Gravitation and Quantizations}, Proceedings of the 1992 Les Houches Summer School, edited by B. Julia and J. Zinn-Justin (North-Holland, Amsterdam, 1994).
\bibitem{Marinov} M. S. Marinov and B. Segev, Report No. quant-ph/0301
\bibitem{Grot} N. Grot, C. Rovelli e R. S. Tate, Phys. Rev. A 54, 6 (1996).
\bibitem{Kumar}N. Kumar, Pramana J. Phys. 25, 363 (1985).
\bibitem{toperator} G. Torres-Vega, Phys. Rev. A 75, 032112 (2007).
\bibitem{time} V. Giovannetti, S. Lloyd, and L. Maccone, arXiv:1504.04215v2 [quant-ph] 4 Jun 2015.
\bibitem{Wooters} D.N. Page and W.K. Wootters, Phys. Rev. D, 27, 2885 (1983).
\bibitem{Parisio1} F. Parisio, Physical Rev. A 84, 062108, (2011).
\bibitem{Parisio2} M. G. Moreno and F. Parisio, Physical Review. A 88, 012118, (2013).


\end{thebibliography}
\end{document}